\documentclass[aip,reprint]{revtex4-1}
\usepackage{graphicx}
\usepackage{multirow}

\draft % marks overfull lines with a black rule on the right

\begin{document}

%\preprint{}

\title{Green's Function Coupled Cluster Simulation of the Near-valence Ionizations of DNA-fragments}
% Force line breaks with \\

\author{Bo Peng}
 \email{peng398@pnnl.gov}
 \affiliation{Advanced Computing, Mathematics, and Data Division, Battelle, Pacific Northwest National Laboratory, K8-91, P.O. Box 999, Richland WA 99352, USA}%Lines break automatically or can be forced with \\
\author{Karol Kowalski}%
 \email{karol.kowalski@pnnl.gov}
\affiliation{ 
William R. Wiley Environmental Molecular Sciences Laboratory, Battelle, 
Pacific Northwest National Laboratory, K8-91, P.O. Box 999, Richland WA 99352, USA%\\This line break forced with \textbackslash\textbackslash
}%

\author{Ajay Panyala}
\author{Sriram Krishnamoorthy}
\affiliation{%
High Performance Computing, Battelle, 
Pacific Northwest National Laboratory, K8-91, P.O. Box 999, Richland WA 99352, USA%\\This line break forced% with \\
}%

\date{\today}% It is always \today, today,

\begin{abstract}
Accurate description of the ionization process in DNA is crucial to the understanding of the DNA damage under exposure to ionizing radiation, and the exploration of the potential application of DNA strands in nano-electronics. In this work, by employing our recently developed Green's function coupled-cluster (GFCC) library on supercomputing facilities, we have studied the spectral functions of several guanine$-$cytosine (G$-$C) base pair structures ([G$-$C]$_n$, $n=1-3$) for the first time in a relatively broad near-valence regime  ([-25.0,-5.0] eV) in the coupled-cluster with singles and doubles (CCSD) level. Our focus is to give a preliminary many-body coupled-cluster understanding and guideline of the vertical ionization energy (VIE), spectral profile, and ionization feature changes of these systems as the system size expands in this near-valence regime. The results show that, 
as the system size expands, even though the lowest VIEs keep decreasing, the changes of spectral function profile and the relative peak positions get unexpectedly smaller.
Further analysis of the ionized states associated with the most intensive peak in the spectral functions reveals non-negligible $|2h,1p\rangle$'s in the ionized wave functions of the considered G$-$C base pair systems. The leading $|2h,1p\rangle$'s associated with the main ionizations from the cytosine part of the G$-$C base pairs feature a transition from the intra-base-pair cytosine $\pi\rightarrow\pi^\ast$ excitation to the inter-base-pair electron excitation as the size of G$-$C base pairs expands, which also indicates the minimum quantum region in the many-body calculations of DNA systems.
\end{abstract}

\maketitle

\section{Introduction}

Accurate description of the electronic structure of the DNA of living organisms is vital for the studies of DNA radiation damage,\cite{lehnert07} DNA redox sensing/labeling,\cite{barton10_891} and charge transport along the double helix in nanoelectronics.\cite{giese02_51,tetsuro13_2616,Dekker01_29} Theoretically, there have been tremendous efforts working towards characterizing the electronic structures of nucleobases, nucleotides, and base pair sequences in terms of ionization energies, electron affinities, and redox potentials (see Reference \citenum{jungwirth15_1209} for a recent overview), in which many computational efforts in recent years have been paid for finding a proper way to describe the vertical ionization energies (VIEs). 

It has been found from extensive density functional theory (DFT) calculations,\cite{jungwirth09_6460,slavicek11_1294,rene14_532,ghosh16_6526,ursula19_2042} as well as Hartree$-$Fock (HF),\cite{saito96_7063,okada09_16384} and quantum mechanical/molecular mechanical (QM/MM) multilevel quantum calculations,\cite{jungwirth13_3766} that the lowest VIEs of DNA fragments in the gas phase greatly depend on the size and the sequence of the DNA fragments. Even though this dependence in the aqueous solution has turned out to be unexpectedly small,\cite{jungwirth09_6460,jungwirth13_3766} recent DFT studies employing implicit solvent models were still unable to witness the convergence of the lowest VIEs for a wide range of DNA fragments considered ($e.g.$ up to three G$-$C base pairs in Reference \citenum{ursula19_2042}). For higher VIEs, the DFT approach with Koopmans-Compliant functionals\cite{marzari16_3948} has been employed to simulate the photoemission spectroscopy of single nucleobases corresponding up to $\sim$20 eV VIEs showing good agreements with the experimental data, while the computation of the ionizations of longer sequence has rarely been reported in the similar theoretical framework. In the meanwhile, it should be realized that the quality of the DFT results heavily depends on the choice of density functionals, and the employed density functionals need to be validated by more accurate and predictable methods to attenuate the large self-interaction energy (SIE) and correct the over-delocalized charge density.\cite{perdew81_5048,perdew96_3865} 

Regarding to the practice of highly accurate and predictable methods in this field, there have been reports in recent years focusing on the studies of the VIEs of small DNA fragments ($e.g.$ single nucleobase or base pair) employing post-Hartree$-$Fock methods such as M{\o}ller-Plesset perturbation theory, coupled-cluster theory, multireference methods, equation-of-motion approaches and Green's function formalism.\cite{serrano06_084302, krylov10_12305, krylov12_2726, krylov10_2001, krylov10_2292, lievin06_9200, jungwirth13_3766, ghandehari13_6027, ortiz06_13350} In particular, the equation of motion coupled cluster method for ionization energy with single and double excitations (EOM-IP-CCSD) method\cite{stanton93_7029} has often been used.\cite{krylov12_2726, krylov10_2001, krylov10_2292} However, for the ionizations of larger DNA fragments, since the dimension of the Hamiltonian matrix in the ($N$$-$1) particle space grows polynomially, the computational cost of these many-body methods becomes quite prohibitive. 

In general, to overcome the computational challenges and leverage the accuracy of high-level many-body methods, the algorithms and approximations to these methods used in describing the electronic structure of large molecular systems need to be carefully designed, and the computational tools (including both hardware and software) need to be systematically optimized to take care of the expensive tensor contractions and communications. Recently, based on the Green's function coupled cluster (GFCC) formalism.\cite{nooijen92_55, meissner93_67, nooijen93_15, nooijen95_1681} we have proposed a highly efficient approach of solving a frequency-dependent linear system to directly compute the frequency-dependent GFCC matrix elements for molecular systems without explicitly knowing the pole structures of similarity transformed Hamiltonians represented in $N\pm1$ electron Hilbert spaces.\cite{kowalski18_4335, kowalski18_214102} Similar methods have been applied in the computation of the spectral functions of simple systems including uniform electron gas,\cite{chan16_235139} light atoms,\cite{matsushita18_034106} heavy metal atoms,\cite{matsushita18_224103} and 1-D periodic systems.\cite{matsushita18_204109} 
By further employing model-order-reduction (MOR) technique in this framework, the approximate GFCC approach is able to compute the spectral function of molecular systems over a broader frequency range at a relatively cheaper cost.\cite{peng19_3185} 

In this communication, with the aid of our recently developed numerical library (specifically designed for many-body calculations)\cite{Mutlu2019} and supercomputing facility, we apply this approximate GFCC approach with singles and doubles ($i.e.$ GFCCSD) to compute the ionizations of three G$-$C base pair structures, [G$-$C]$_n$ ($n=1-3$). This work aims to understand in a many-body coupled-cluster way the ionizations of the relatively longer G$-$C base pair sequences in the near-valence regime and their features. In particular, we try to find out (i) how the ionizations in this regime and their features change as the system size expands, and (ii) how the many-body coupled cluster description would be different from the single-particle picture and lead us to a more generalized near-valence ionization picture of longer DNA sequence. 

\section{Computational Methods}

For an overview of the GFCC approach and its approximation used in this work, we refer the readers to References \citenum{kowalski18_4335}, \citenum{kowalski18_214102}, and \citenum{peng19_3185}. Briefly, it is an approach to compute the matrix element of the analytical frequency-dependent one-particle coupled-cluster Green's function of an $N$-electron system. The expression of its retarded part that is associated with ionization can be written as 
\begin{eqnarray}
G^R_{pq}(\omega) = 
&&\langle\Phi|(1+\Lambda) \bar{a_q^{\dagger}} (\omega+\bar{H}_N- \text{i} \eta)^{-1} \bar{a}_p |\Phi\rangle. 
\label{gfxn1}
\end{eqnarray}
In this equation, $|\Phi\rangle$ is the reference wave function, $a_p$ ($a_p^\dagger$) operator is the annihilation (creation) operator for the electron in the $p$-th spin-orbital, and $\bar{H}_N$ is similarity transformed Hamiltonian $\bar{H}$ ($\bar{H} = e^{-T} H ~e^{T}$) in a normal product form. The cluster operator $T$ and the de-excitation operator $\Lambda$ are obtained by solving the conventional coupled cluster equations. The evaluation of Eqn. (\ref{gfxn1}) is addressed by solving a linear system for an auxiliary excitation amplitude that corresponds to the ionized state, which has been detailed in the our previous work.\cite{kowalski18_4335, kowalski18_214102} The spectral function is then given by the trace of the imaginary part of the retarded GFCC matrix,
$A(\omega) = - \frac {1} {\pi} \text{Tr} \left[ \Im\left({\bf G}^{\text{R}}(\omega) \right) \right]$.

The conventional GFCC calculation needs to compute the Green's function matrix elements for every single frequency of interest, and is thus bounded by the number of frequency points, $N_{\omega}$, constituting a sizable pre-factor to the already large complexity of the calculation. Take the GFCCSD calculation for an example, the total cost could mount up to $\sim$$\mathcal{O}(N_{\omega}N^6)$ with $N$ being the number of basis functions representing the system size. By employing the MOR techniques to project the conventional GFCC approach to a manageable subspace approach, and by interpolating and extrapolating more frequencies to some extent, we were able to significantly reduce $N_{\omega}$, and nicely reproduce the GFCC spectral functions obtained from the conventional approach for small and medium size molecular systems in both the core and near-valence regimes.\cite{peng19_3185} More details of our approximate subspace GFCCSD approach used in the present work has been demonstrated in the supplementary material. All the geometries used in this calculation were obtained from Reference \citenum{ursula19_2042}, and were optimized at the B3LYP/6-31++G(d) level for the neutral systems under an implicit solvation treatment where a Polarizable Continuum Model (PCM) has been applied for water.\cite{miertus81_117, amovilli98_227, corni02_5697}

\section{Results and Discussion}

For the present study, we first test the accuracy of the approximate GFCCSD approach by computing the spectral functions of single cytosine and guanine bases in the regime of [-25.0,-5.0] eV, and comparing the obtained GFCCSD spectral functions with the previously reported single-particle DFT and many-body ADC(3) results.  Based on previous basis set benchmarks,\cite{ursula19_2042, ghosh16_6526} and given the wide range of the system sizes considered, the 6-31++G(d) basis set has been chosen for all the GFCCSD calculations in the present study (the difference between the double-$\zeta$ and triple-$\zeta$ basis sets were found to be $\le$0.1 eV from our previous GFCCSD study on the valence and core ionizations of some small and medium size molecules, see References \citenum{kowalski18_4335} and \citenum{peng19_3185}).
As can be seen from Figure S1, between the single-particle KS-DFT results and the many-body ADC(3) and GFCCSD results, qualitative agreements can be reached for the overall spectral function profiles except that the KS-DFT exhibits $\sim$2 eV red shift at high energy regime ($<$-15.0 eV) and different degeneracies at some low energies. 
Between the ADC(3) and the GFCCSD results, due to the inclusion of the $|2h,1p\rangle$ configurations in both methods, excellent agreements can be observed in terms of peak positions, peak heights, and entire profiles. The minor difference comes from the slight shift of some interior peaks (see the cytosine spectral function at $\sim$-15 eV, and the guanine spectral function at $\sim$-13 eV). 
Since it is experimentally hard to extract and establish the ionizations of the single cytosine or guanine tautomer used in the present study, the comparison between our GFCCSD results and the experimental data can only be made indirectly. For the present guanine molecule, the experimental first VIE has been estimated at different high level theories to be roughly 7.85$\sim$8.15 eV.\cite{hush75_11, dougherty78_379, serrano06_084302, ahmed07_7562, ahmed09_4829} Thus, 
the computed first VIE at the GFCCSD/6-31++G(d) level ($\sim$7.78 eV) is 0.08$\sim$0.37 eV lower than the estimated experimental value, which is in line with the typical error of VIEs computed by the EOM-CCSD method.\cite{bartlett07_291}

\begin{figure}
\includegraphics[width=0.45\textwidth]{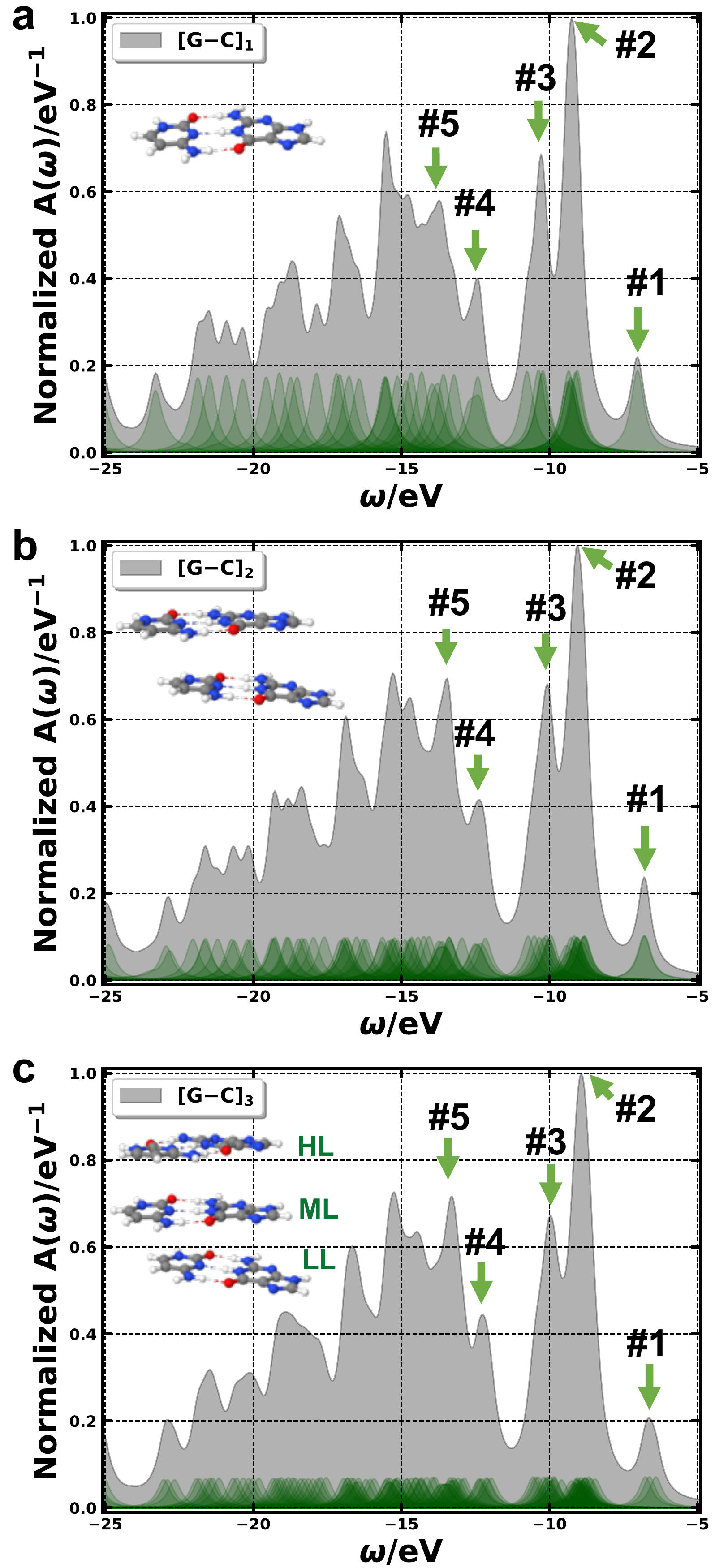}
\caption{Spectral functions of G$-$C base pair systems, $i.e.$ (a) [G$-$C]$_1$, (b) [G$-$C]$_2$, and (c) [G$-$C]$_3$, in the near-valence regime computed at the GFCCSD/6-31++G(d) level. The number of basis functions used for [G$-$C]$_1$, [G$-$C]$_2$, and [G$-$C]$_3$ are 391, 783, and 1173, respectively. The spectral functions are shown in the grey, and the five main peaks in the lower energy regime ([-14.0,-5.0] eV) in each spectral function are marked by green arrows. The frequency interval $\Delta\omega$ = 0.05 eV and the broadening factor $\eta$ = 0.27 eV. The contributions to the spectral functions ($i.e.$ the imaginary part of the diagonal matrix elements of the retarded Green’s function) are shown in green. The VIEs that correspond to the absolute values of the main peak positions are given in the supporting information.  
For three-layer [G$-$C]$_3$, the high layer (HL), middle layer (ML), and low layer (LL) G$-$C base pairs have also been labelled.
\label{fig2}}
\end{figure}

After validating our GFCCSD approach for the single cytosine and guanine bases, we then apply the GFCCSD approach to compute the spectral functions of three G$-$C base pair structures, [G$-$C]$_n$ ($n=1-3$), and the results are shown in Figure \ref{fig2}. At the first glance, the entire profiles of the normalized spectral functions of all the G$-$C base pair structures are very similar to each other in the considered valence regime. For the energy regime of [-17.0,-5.0] eV, the change of the spectral profiles between [G$-$C]$_2$ and [G$-$C]$_3$ becomes unexpectedly smaller than that between [G$-$C]$_1$ and [G$-$C]$_2$ (especially when considering the relative height of the peak \#5 w.r.t. the surrounding peaks). For higher energy regime, as the system size expands, a transition from discretized peaks (Figure \ref{fig2}a) to band-like distribution (Figure \ref{fig2}c) can be observed, while the entire spectral profiles are still roughly consistent for all the considered G$-$C base pair structures.

\begin{figure}
\includegraphics[width=0.45\textwidth]{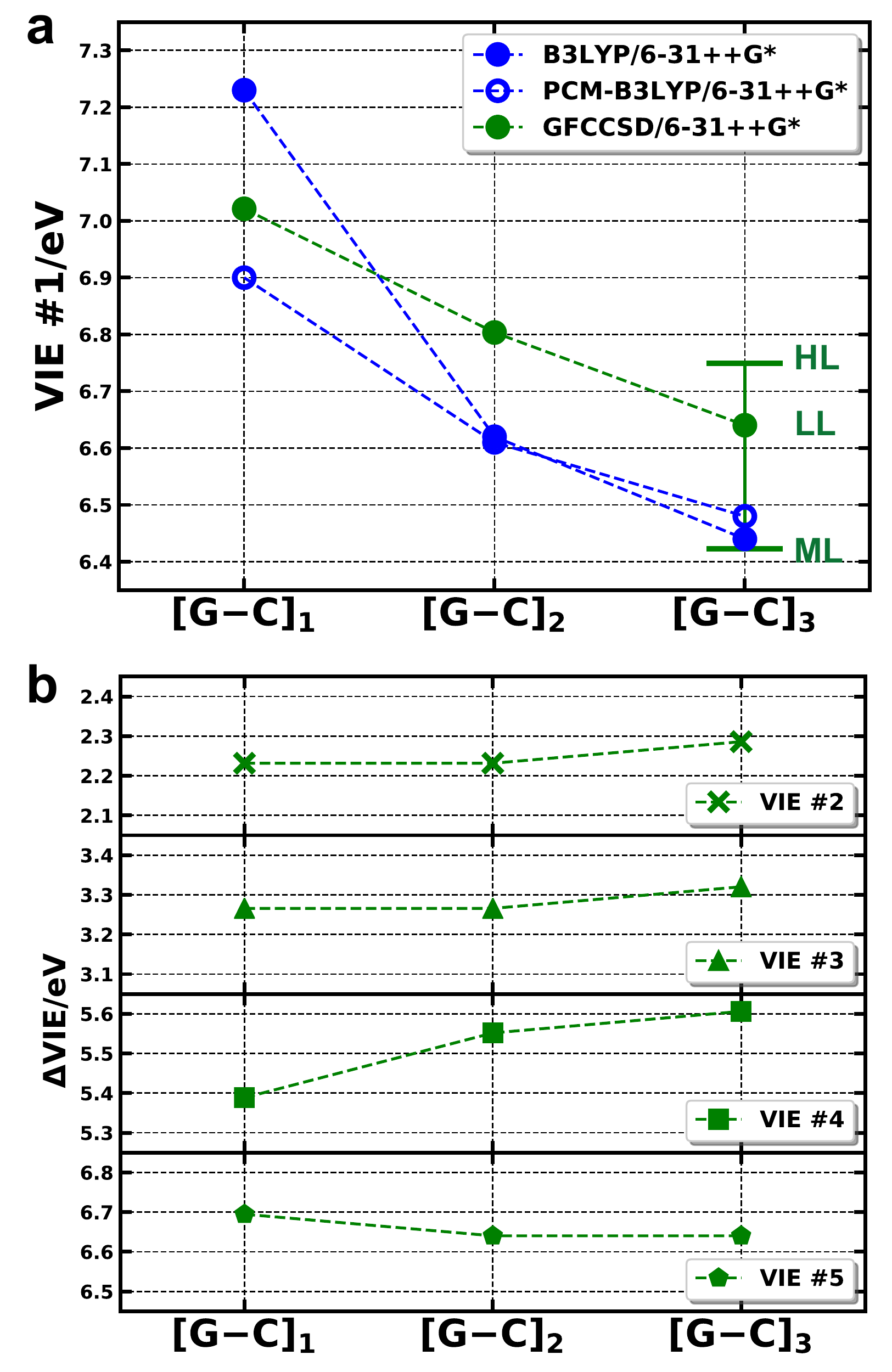}
\caption{The changes of five main VIEs in [-14.0,-5.0] eV as functions of the G$-$C base pair system size. (a) shows the change of the VIE \#1 of all the considered G$-$C base pair structures at different levels of theory, and (b) shows the changes of the relative positions of the other four VIEs w.r.t. the VIE \#1 (denoted as $\Delta$VIEs) from the present GFCCSD calculations. Here, the VIEs are read from the GFCCSD spectral functions in Figure \ref{fig2}. In (a), the B3LYP results were obtained from Reference \citenum{ursula19_2042}. The first VIEs of each G$-$C layer in [G$-$C]$_3$ have also been marked in (a).
\label{fig3}}
\end{figure}

To have a close look at how the peak positions (or the VIEs) change as the system size expands, we select the five main low-energy peaks in [-14.0,-5.0] eV from the spectral function of each G$-$C base pair system, and plot them as functions of the system size in Figure \ref{fig3} (the values of all the main VIEs in the near-valence regime are given in the supplementary material). In Figure \ref{fig3}a, the first main VIE (VIE \#1) computed by the GFCCSD approach in this work is reduced from $\sim$7.02 eV in [G$-$C]$_1$ to $\sim$6.64 eV in [G$-$C]$_3$, showing same tendency as the gas phase B3LYP results based on the same geometries and base set.\cite{ursula19_2042}  Further comparison shows the B3LYP result is $\sim$0.2 eV higher than the GFCCSD result for [G$-$C]$_1$, which is close to the deviation reported in another work\cite{ghosh16_6526} where the gas phase EOM-IP-CCSD result of a similar G$-$C single base pair was $\sim$0.15 eV lower than the DFT result using $\omega$B97x-D density functional. As the system expands to [G$-$C]$_2$ and [G$-$C]$_3$, the B3LYP results then become consistently $\sim$0.2 eV lower than the GFCCSD results. 
Remarkably, the peak \#1 in Figure \ref{fig2}c is composed of three Green's function peaks (green curves), in which the rightmost peak gives the lowest VIE of [G$-$C]$_3$ that corresponds to the ionization of the middle G$-$C layer in [G$-$C]$_3$, and is $\sim$0.01 eV lower than the B3LYP result (see Figure \ref{fig3}a).
In the Reference \citenum{ursula19_2042}, solvent effect has also been considered in the computation of the first VIEs of the G$-$C base pairs by employing the non-equilibrium PCM solvent model. As can be seen, after including the non-equilibrium PCM solvent model in the B3LYP calculations, the first VIE was dropped by $\sim$0.3 eV for [G$-$C]$_1$, while the VIE shifts for [G$-$C]$_2$ and [G$-$C]$_3$ were $<$0.05 eV. 
One may expect similar solvent effect on the many-body values.
For the G$-$C base pairs ($i.e.$ [G$-$C]$_1$ and [G$-$C]$_2$) in Dickerson dodecamer, the QM/MM ($i.e.$ MP2-aug-cc-PVDZ/PM6) calculations under implicit solvent treatment have shown to bring down the gas phase first VIE by 0.36 eV for  [G$-$C]$_1$ and 0.07 eV for [G$-$C]$_2$, respectively,\cite{jungwirth13_3766,jungwirth15_1209} which were very close to the deviations between the B3LYP and PCM-B3LYP results. 
Nevertheless, it is worth mentioning that the proper inclusion of the solvent effect in many-body calculation is still challenging. Extensive evaluation of the EOM-CCSD method combined with the PCM for the calculation of electronic excitation energies of solvated molecules has shown that the EOM-CCSD-PCM approach consistently overestimated experimental results by 0.4$\sim$0.5 eV, which is slightly larger than the expected EOM-CCSD error in vacuo (about 0.1$\sim$0.3 eV).\cite{caricato17_117} 

Different from the change of the first main VIEs, 
the change of higher VIEs w.r.t. the first VIEs, $\Delta$VIEs, computed from our GFCCSD approach are much smaller. As shown in Figure \ref{fig3}b, from [G$-$C]$_1$ to [G$-$C]$_3$, the $\Delta$VIE changes for the other four main VIEs are mostly $\le$0.05 eV. Regarding to the ionized states, 
the configuration analysis in Figure S2
shows larger $|2h,1p\rangle$ contribution to the ionized state as the VIEs go higher. More importantly, all the considered G$-$C base pair systems exhibit consistent $|2h,1p\rangle$ contributions to the ionized states, $P_{X_2}(\omega)$, for the VIEs up to $\sim$17 eV. For higher ionizations, significant difference in the $P_{X_2}(\omega)$ can be observed between the [G$-$C]$_1$ and [G$-$C]$_3$, and the difference generally becomes larger (from $\sim$5\% to $\sim$15\%) as the system size expands. Between the [G$-$C]$_2$ and [G$-$C]$_3$, the difference in the $P_{X_2}(\omega)$ becomes much smaller for most part of the regime except that the difference climbs up to 0.11 at the VIE of $\sim$23 eV. 

\begin{table}
\resizebox{0.5\textwidth}{!}{%
\begin{tabular}{lllrlclllcl}
\hline \hline
 & \multirow{2}{*}{Fragments}                 			&  & \multicolumn{1}{l}{\multirow{2}{*}{VIE/eV}} 	&  & \multicolumn{1}{l}{\multirow{2}{*}{\begin{tabular}[c]{@{}l@{}}Ionized \\ orbital\end{tabular}}} 	&  &  \multicolumn{1}{c}{\multirow{2}{*}{Major $|2h,1p\rangle$'s}} & 
 &  \multicolumn{1}{l}{\multirow{2}{*}{\begin{tabular}[c]{@{}l@{}}Weight of \\ $|2h,1p\rangle$'s\end{tabular}}}   & \\
 &                                            					&  &                             				&  &                                                                          						&  &                                                                                	&  & &  \\
 &                                            					&  &                             				&  &                                                                            						&  &                                                                                    	&  & & \\  
 \hline
 &                                            					&  &                            				&  &                                                                             						&  &                                                                            		&  & & \\
 & \multicolumn{1}{c}{\multirow{2}{*}{[G$-$C]$_1$}}   &  &  \multicolumn{1}{c}{\multirow{2}{*}{9.25}}                   &  & \multicolumn{1}{c}{\multirow{2}{*}{$H_{-1}$}}                                                                  		&  & $H_{-1}^{-1}H_{-11}^{-1}L_{+5}$, $H_{-1}^{-1}H_{-11}^{-1}L_{+13}$, & &  \multicolumn{1}{c}{\multirow{2}{*}{7.40\%}} & \\
 &		   								&  &                                             					&  &                                                                                        								&  & $H_{-1}^{-1}H_{-11}^{-1}L_{+22}$, $H_{-1}^{-1}H_{-11}^{-1}L_{+28}$  & & &  \\. 
 &		   								&  &                                             					&  &                                                                                        								&  & 											& & &  \\
 & \multicolumn{1}{c}{\multirow{2}{*}{[G$-$C]$_2$}} 	&  & \multicolumn{1}{c}{\multirow{2}{*}{9.04}}                	&  &  \multicolumn{1}{c}{\multirow{2}{*}{$H_{-2}$}}                                     						&  & $H_{-2}^{-1}H_{-22}^{-1}L_{+40}$, $H_{-2}^{-1}H_{-22}^{-1}L_{+11}$,  & & \multicolumn{1}{c}{\multirow{2}{*}{7.69\%}} & \\
 &		   								&  &                                             					&  &                                                                                        								&  & $H_{-2}^{-1}H_{-22}^{-1}L_{+12}$, $H_{-2}^{-1}H_{-22}^{-1}L_{+44}$  & & &  \\ 
 &		   								&  &                                             					&  &                                                                                        								&  & 											& & &  \\
 & \multicolumn{1}{c}{\multirow{2}{*}{[G$-$C]$_3$}}  	&  &  \multicolumn{1}{c}{\multirow{2}{*}{8.93}}                   	&  &  \multicolumn{1}{c}{\multirow{2}{*}{$H_{-3}$}}                                                                  		&  & $H_{-3}^{-1}H_{-33}^{-1}L_{+58}$, $H_{-3}^{-1}H_{-33}^{-1}L_{+16}$, & & \multicolumn{1}{c}{\multirow{2}{*}{7.84\%}} &  \\
 &                                            					&  &                              							&  &                                                                             										&  & $H_{-3}^{-1}H_{-33}^{-1}L_{+51}$      	& & &  \\ 
 &                                            					&  &                              							&  &                                                                           										&  &                                                                            		& & &  \\
\hline
\end{tabular}
}
\caption{The major $|2h,1p\rangle$'s and their total weight in the ionized states for the specified ionized orbitals in the considered G$-$C base pair systems. In the table, `H' and `L' denote the highest occupied molecular orbital (HOMO) and the lowest unoccupied molecular orbital (LUMO), respectively, and the subscripts denote the offsets from HOMO or LUMO ($e.g.$, $H_{-1}$ and $L_{+5}$ refer to HOMO-1 and LUMO+5 orbitals).  The superscript `-1' denotes that one electron is removed from the corresponding orbitals.
\label{feature}}
\end{table}

To have a preliminary picture of the feature of the near-valence ionizations, the ionized states corresponding to the most intensive main ionization at $\omega\sim-9$ eV have been further analyzed. 
As shown in Table \ref{feature}, for these ionized states, as the system size expands, we examine the ionizations associated with the  HOMO-1 of [G$-$C]$_1$, the HOMO-2 of [G$-$C]$_2$, and the HOMO-3 of [G$-$C]$_3$, respectively. The analysis of the corresponding ionized states shows that  there are about $\sim$8\% $|2h,1p\rangle$ in these ionized states, and the $|2h,1p\rangle$ contribution slightly increases as the system size expands. The molecular orbitals involved in the leading $|2h,1p\rangle$ configurations and their orbital energies at the Hartree-Fock level are shown in Figure S3. As can be seen, different from the first VIEs, where the ionization is mostly on the guanine part, the ionizations of the  HOMO-1 of [G$-$C]$_1$, the HOMO-2 of [G$-$C]$_2$, and the HOMO-3 of [G$-$C]$_3$ are mostly on the cytosine part of the base pairs. Also, as the base pair number increases, 
the leading $|2h,1p\rangle$'s feature a transition from the cytosine $\pi\rightarrow\pi^\ast$ intra-base-pair electron excitation in [G$-$C]$_1$ to an inter-base-pair electron excitation in [G$-$C]$_2$ and [G$-$C]$_3$. 
Besides, the orbital energies ($\epsilon_{orb.}$) of the occupied MOs involved in these ($2h,1p$) interactions only vary slightly. In contrast, the variation of the $\epsilon_{orb.}$'s of the virtual MOs first exhibits a dramatical increase from 1.82 eV (LUMO+5 of [G$-$C]$_1$) to 4.41 eV (LUMO+40 of [G$-$C]$_2$), but then 
a slight shift
from [G$-$C]$_2$ to [G$-$C]$_3$ with the difference between the $\epsilon_{L+40}$ in  [G$-$C]$_2$ and $\epsilon_{L+58}$ [G$-$C]$_3$ being $<$0.2 eV. 
Similar inter-base-pair $\pi\rightarrow\pi^\ast$ excitations also accompany the ionization of the guanine part associated with the first main VIEs. 
The ionization analysis can also be done by employing an $\omega$-dependent orbital basis ($e.g.$ Dyson orbitals) in our approximate GFCC calculation. Practically, the $\omega$-dependent Dyson orbital used in our GFCC calculation can be generated from cheaper perturbative calculations (e.g. a second-order self-energy ($\mathbf{\Sigma}^{(2)}(\omega)$) calculation in our implementation). A more rigorous analysis of the (2h,1p) character of the ionization requires the calculation of two-particle GFCC matrix. Relevant analysis and discussion in the GFCC framework will be presented in our future work.
It is worth noting that the ($2h,1p$) electron interaction between the guanine (or cystosine) units in the stacked G$-$C structures can not be described by a single particle picture with or without the dielectric environment, neither by the QM/MM scheme where only one base pair is fully quantum mechanically treated.\cite{jungwirth15_1209}

\section{Conclusion}
As shown in this work, by employing our recently developed parallel GFCCSD implementation and supercomputing facility, we have for the first time been able to study the ionizations of several G$-$C base pair structures, [G$-$C]$_n$ ($n=1-3$), in a relatively broader near-valence regime of [-25.0, -5.0] eV. The GFCCSD spectral function profiles of the single cytosine and guanine base have shown excellent agreement with other many-body results. For larger G$-$C base pair systems, similar to the previous DFT results, the first main VIEs {keep decreasing} as the system size expands. Nevertheless, the changes of entire spectral profiles in this near-valence regime get unexpectedly smaller. Further analysis on the leading $|2h,1p\rangle$'s at the second main VIEs (where the ionizations are the most intensive) has featured a clear transition from intra-base-pair election excitation in single G$-$C base pair to inter-base-pair excitation in stacked larger G$-$C base pair structure. Such a ($2h,1p$) interaction can not be observed from single particle calculations, nor from many-body simulations of single base pair structure. Our study thus exhibits the importance of many-body approach for accurately mimicking the electronic structure of the DNA systems, and suggests a minimum quantum chemical region containing at least two base pairs for the many-body calculations of the DNA systems.

\section{Supplementary Material}
The supplementary material includes the computational details, the GFCC spectral functions of cytosine and guanine, the main VIE values of [G$-$C]$_n$ ($n=1-3$) and
the weighted $|2h,1p\rangle$ contributions to the ionized states that correspond to these VIEs in the regime of [-25.0,-5.0] eV, and the MOs  involved in the major $|2h,1p\rangle$'s in Table \ref{feature} and their corresponding orbital energies.

\begin{acknowledgments}
The development of the GFCC library was supported by the Center for Scalable, Predictive methods for Excitation and Correlated phenomena (SPEC), which is funded by the U.S. Department of Energy (DOE), Office of Science, Office of Basic Energy Sciences, the Division of Chemical Sciences, Geosciences, and Biosciences. KK also acknowledges the support from the capability development project at the Environmental Molecular Sciences Laboratory (EMSL) in the Pacific Northwest National Laboratory (PNNL) for the development of the second-order self-energy module in the NWChem quantum chemical software suites. PNNL is operated for the U.S. Department of Energy by the Battelle Memorial Institute under Contract DE-AC06-76RLO-1830. This research used resources of the Oak Ridge Leadership Summit Computing Facility, which is a DOE Office of Science User Facility supported under Contract DE-AC05-00OR22725. 
\end{acknowledgments}

\bibliography{gfcc.bib}

\end{document}